\documentclass[11pt]{article}
\usepackage[utf8]{inputenc}
\usepackage[titletoc]{appendix}
\usepackage[usestackEOL]{stackengine}
\usepackage{lipsum}
\edef\tmp{\the\baselineskip}
\setstackgap{L}{\tmp}
\usepackage{amsmath,amssymb,tabularx,booktabs}
\usepackage{extarrows} 
\usepackage{float}
\begin{document}
\title{Sea contribution to the charge radii and quadrupole moment of $J^P=\frac{1}{2}^+, \frac{3}{2}^+$ baryons}
\author{Preeti Bhall*, Meenakshi Batra, A. Upadhyay}
\date{School of Physics and Material Science, \\
Thapar Institute of Engineering and Technology, Patiala, Punjab-147004\\
Dev Samaj College for Women, Sector 45, Chandigarh-160047\\
\today}
\maketitle
\thanks{Email: preetibhall@gmail.com},
\thanks{alka.iisc@gmail.com}
\begin{abstract}
An operator formalism is used on the wavefunction of baryons to compute their charge radii and quadrupole moments. Total anti-symmetric wavefunction in spin, color and flavor space is framed for $J^P=\frac{1}{2}^+$ nucleons and $J^P=\frac{3}{2}^+$ hyperons. To understand the importance of sea, statistical model is used in conjugation with the detailed balance principle. Within the statistical approach, the importance of sea with quarks and gluons are studied using the relevant probabilities that are associated with spin, flavor, and color space. 
The present work also focuses on individual contributions of valence and sea which contains terms of scalar, vector and tensor sea. The obtained results are in agreement with available  theories and few experimental outcomes. Our computed results may provide important information for upcoming experimental findings.
\end{abstract}
\section{Introduction}
The main goal for particle physicists is to understand the structure of hadrons. The study of electromagnetic properties of baryons over the last two decades enhances our understanding of the internal structure of baryons and gives new directions about the properties of matter. Various discoveries of proton suggest that quarks-antiquarks carry 30$\%$ of the total proton spin. Further, it has been predicted \cite{01} that the orbital angular momentum of quarks and gluons also make a significant contribution to the proton's spin and other static properties \cite{02} of the baryons like magnetic moment, masses, etc. It is also confirmed that there is a significant effect from the strange quarks \cite{03,04} which are assumed to be present in the sea. Usually, the 'sea' is a combination of virtual gluons and quark-antiquark pairs. It may contain light quarks as well as heavy ones and plays a vital role in determining the hadronic structure. Therefore, to go into deeper details of the structure of baryons, it becomes mandatory to investigate their low energy properties. The most common properties to study the structure of hadrons are magnetic moments, masses, semi-leptonic decays, charge radii, quadrupole moment and spin distribution etc. Although, Quantum Chromodynamics(QCD) is the fundamental theory of strong interactions of hadrons but presently these properties can not be predicted directly from QCD. The study of hadronic structure is complicated due to the non-perturbative interaction between the fundamental constituents (i.e. quarks and gluons). To get an interior image of the baryons, the theorist depends upon phenomenological models and effective theories. Over the past years \cite{05,06,07}, we have witnessed remarkable discoveries in the area of phenomenology. The charge radii and quadrupole moments are important observables of hadrons that contain information about their quark-gluon structure. These are directly related to the spatial charge and current distribution in baryons. Various scattering experiments \cite{08} have been made measurements for the charge radii of the proton, neutron and $\Sigma^{-}$ giving values $r_p$ = 0.8409 $\pm$ 0.0004 fm, $r_{n}^2$ = -0.115 $\pm$ 0.0017 fm$^2$ and $\Sigma^{-}$ = 0.78 $\pm$ 0.10 fm listed in PDG \cite{07}. On reviewing the previous literature, many theoretical models like the quark model \cite{09,10,11}, light cone QCD sum rules \cite{12,13}, the Skyrme model with bound-state approach \cite{14},  QCD parametrization method \cite{15,16}, the 1/N$_c$ expansion method\cite{17,18}, lattice QCD \cite{19,20} and chiral constituent quark model ($\chi$QM) \cite{21,22,23} analysed the various properties of baryons such as magnetic moment, charge radii, octupole moments, quadrupole moments etc. By using the QCD sum rule approach in ref. \cite{12,13}, the author estimated the magnetic dipole, electric quadrupole, and magnetic octupole moments of $\Delta$-baryons. In ref. \cite{17,18}, the authors predicted the charge radii and quadrupole moment of decuplet baryons by 1/$N_c$ expansion method.  In ref. \cite{19,20}, the charge radii of decuplet baryons have been studied in lattice QCD. In ref. \cite{21, 22, 23}, the charge radii, quadrupole moment of $(\frac{1}{2})^{+}$ octet, $(\frac{3}{2})^{+}$ decuplet baryons and spin $(\frac{3}{2})^{+}\rightarrow(\frac{1}{2})^{+}$ transitions are calculated in the framework of $\chi$CQM using general parameterization (GP) method. An enormous amount of experimental information \cite{07} is available now, which caters to the need of detailed comparisons between theory and experimental data. Also, the decuplet to octet transition moment plays an important role in studying the internal structure of the spin. 
The quadrupole transition moment$(\Delta^+\rightarrow N)$ measured by LEGS \cite{24} and Mainz \cite{25} collaborations $(-0.018\pm0.009\pm0.034$ fm$^2$ and $-0.0846 \pm 0.0033$ fm$^2)$ respectively \cite{26} brings to the conclusion that the nucleon and $\Delta^+$ baryon are intrinsically deformed.\\
There are many theoretical approaches in the literature to study electromagnetic properties. But the results from different theoretical models are not consistent with each other. In order to have a clear understanding, we study these electromagnetic properties (quadrupole moment and charge radii) in the framework of the statistical model with the principle of detailed balance. In this model, hadrons are assumed to consist of valence quarks which are associated with a virtual dynamic sea where quark-antiquark pairs are multi-connected through gluons. Singh \& Upadhyay et al. \cite{27} used the statistical model to find the quark contribution to the spin of the nucleons. The statistical model with the inclusion of 'sea quarks' successfully explains the low energy hadronic properties like masses \cite{28}, spin distribution \cite{29}, magnetic moments \cite{30} of octet and decuplet baryons, and importance of sea contribution to nucleons \cite{31}. As this approach has been proven to be quite effective in the calculation of static properties of baryons. So, we extended the statistical model along with the principle of detailed balance by computing the quadrupole moments and charge radii of the ground state nucleons and $\Delta$-hyperons. \\ 
This paper is organised in the following order: In sec. 2, a brief overview of octet and decuplet baryons wavefunction with sea components is given. Sec. 3, presents the explanation of quadrupole moment and charge radii operator. In Sec. 4, the statistical model with the detailed balance principle is discussed. The importance of statistical parameters to the static properties of baryons is also explained. In Sec. 5, we discuss the calculated charge radii and quadrupole moment, and compare the computed values to the recent available experimental and theoretical data. The conclusion is given in the sec. 6.
\section{ Preliminaries}
The naive valence picture of the hadron structure is considered as a first-order approximation to the real system. With the extension of the naive quark model, baryons are assumed a composite system of valence quark (qqq) and a 'sea', where sea contains gluons and virtual quark-antiquark pairs \cite{32,33}. Sea is characterized by its total quantum number consistent with the quantum number of valence quarks. 
The valence quark wavefunction of the baryon \cite{34} is written as: \\      \centerline{$\Psi = \Phi (|\phi> |\xi> |\chi> |\rho>)$}\\
where $|\phi>$, $|\xi>$, $|\chi>$ and $|\rho>$ represent the flavor, spin, color and space $q^{3}$ wave functions respectively. Here, the spatial part $|\rho>$ is symmetric under the exchange of any two quarks. The other three parts flavor-spin-color must be anti-symmetric in order to maintain the overall anti-symmetrization of the baryons. The three valence quarks each having spin-$\frac{1}{2}$ will lead to the possibility of $J^P=\frac{1}{2}^+, \frac{3}{2}^+$ represented as:\\

\centerline{qqq: $\frac{1}{2}\bigotimes\frac{1}{2}\bigotimes\frac{1}{2} =2(\frac{1}{2}) \bigoplus \frac{3}{2}$}
and being the ground state particles, these quarks (fermions) have positive parity which is given by $(-1)^l$.  We take into consideration a flavorless sea having spin (0, 1, 2) and color (1, 8, $\bar {10}$).  Let $H_{0,1,2}$ and $G_{1,8,\bar{10}}$ represent the spin and color sea wave functions respectively which satisfy the condition: {$<H_i|H_j> = \delta_{ij}$, $<G_k|G_l> = \delta_{kl}$}. The 'sea' is supposed to be comprised of a gluon or a $q\bar q$ pair, or more complicated, with multi-gluon state (gg, ggg), multi-$q\bar q$ pairs or gluon(s) plus ($q\bar q$) pairs. The spin (H) and color (G) combination of sea having two gluons, $q\bar q$ or multi $q \bar q$ pairs can be present as: \\

\makebox[\textwidth]{ \textbf{Spin}: gg: $ 1\bigotimes1 = 0_s\bigoplus 1_a\bigoplus 2_S$}\\
\makebox[\textwidth]{$q\bar qq\bar q : (\frac{1}{2}\bigotimes \frac{1}{2})\bigotimes(\frac{1}{2}\bigotimes \frac{1}{2}) = (0_a\bigoplus1_s)\bigotimes(0_a\bigoplus1_s)$}\\
\makebox[\textwidth]{\hspace{2cm} = $2(0_s)\bigoplus 1_s\bigoplus2(1_a)\bigoplus2_s$}\\

\makebox[\textwidth]{\textbf{Color}: gg: $8\bigotimes 8 = 1_s\bigoplus 8_s\bigoplus 8_a\bigoplus 10_a\bigoplus \bar {10_a}\bigoplus 27_s$}\\
\makebox[\textwidth]{$q\bar qq\bar q : (3\bigotimes \bar 3)\bigotimes(3\bigotimes \bar 3) = (1_a\bigoplus8_s)\bigotimes(1_a\bigoplus8_s)$}\\
\makebox[\textwidth]{\hspace{2cm} = $2(1_s)\bigoplus2(8_s)\bigoplus2(8_a)\bigoplus10_s\bigoplus \bar {10_s} \bigoplus 27_s$}\\

Similar treatment is extended for multiple gluon cases upto three. Subscripts \textbf{\textit{a}} and \textbf{\textit{s}} denote the anti-symmetry and symmetry of the combined state. The total flavor-spin-color wave function of baryon octet (decuplet) of three valence quarks with the inclusion of sea components can be represented as:\\
\begin{equation}
\begin{aligned}
  |\Phi_{1/2}^{(\uparrow)}> = & \frac{1}{N} [{\Phi_{1}}^{(\frac{1}{2} \uparrow)}H_{0}G_{1} + a_{8} ({\Phi_{8}}^{(\frac{1}{2})}\otimes H_0)^{\uparrow}G_8 + a_{10} {\Phi_{10}}^{(\frac{1}{2}\uparrow)}H_{0}G_{\bar {10}} +\\& b_1({\Phi_1}^{(\frac{1}{2})} \otimes H_1)^{\uparrow} G_1+b_8 ({\Phi_8}^{(\frac{1}{2})} \otimes H_1) ^{\uparrow} G_8 + b_{10} ({\Phi_{10}}^{(\frac{1}{2})} \otimes H_1)^{\uparrow} G_{\bar {10}}\\& + c_8 ({\Phi_8}^{(\frac{3}{2})} \otimes H_1)^{\uparrow} G_8 + d_8 ({\Phi_8}^{(\frac{3}{2})} \otimes H_2)^{\uparrow} G_8
\end{aligned}
\end{equation}
 where $N^{2} = 1+ a_{8}^{2} +a_{10}^{2} +b_{1}^{2} +b_{8}^{2} + b_{10}^{2} + c_{8}^{2} + d_{8}^{2}$\\
For Decuplet baryons \cite{30}:
\begin{equation}
    \begin{aligned}
  |\Phi_{3/2}^{(\uparrow)}> =& \frac{1}{N} [a_0{\Phi_{1}}^{(\frac{3}{2} \uparrow)}H_{0}G_{1} + b_1({\Phi_1}^{(\frac{3}{2})} \otimes H_1)^{\uparrow} G_1 + b_8 ({\Phi_8}^{(\frac{1}{2})} \otimes H_1) ^{\uparrow} G_8 +\\&d_1 ({\Phi_1}^{(\frac{3}{2})} \otimes H_2)^{\uparrow} G_1 + d_8 ({\Phi_8}^{(\frac{1}{2})} \otimes H_2)^{\uparrow} G_8    
    \end{aligned}
\end{equation}
where $N^{2} = a_{0}^{2} +b_{1}^{2} +b_{8}^{2} + d_{1}^{2} + d_{8}^{2}$\\
Here N is the normalisation constant.  
The symbol ‘$\Phi$’ contains the spin and color of valence quarks. For e.g.- $\Phi_{1}^{(\frac{1}{2} \uparrow)}$ represent the spin-$\frac{1}{2}$ and color singlet state of valence quarks. Similarly, $\Phi_{8}^{(\frac{1}{2} \uparrow)}$ specifies the spin-$\frac{1}{2}$ and color octet of valence quarks. So, the total wavefunction consists of various combination of valence and sea part can be written as: ${\Phi_{1}}^{(\frac{1}{2} \uparrow)}H_{0}G_{1}, ({\Phi_{8}}^{(\frac{1}{2})}\otimes H_0)^{\uparrow}G_8, {\Phi_{10}}^{(\frac{1}{2}\uparrow)}H_{0}G_{\bar {10}}$. The terms $(\Phi_1^{(\frac{1}{2})} \otimes H_1)^\uparrow$, $(\Phi_8^{(\frac{1}{2})} \otimes H_1)^\uparrow$, $(\Phi_1^{(\frac{3}{2})} \otimes H_1)^\uparrow$, etc. written with suitable C.G. coefficients by considering the symmetry property of the component wave function. The sea with spin 0, 1, and 2 represents scalar, vector and tensor sea respectively. 
In octet wave function, the coefficients $a_8, a_{10}$ come from the scalar sea whereas the coefficients $b_1, b_8, b_{10}, c_8$ and $d_8$ are from the vector and tensor sea respectively. Similarly, in the decuplet wavefunction, the coefficient $a_0$ comes with the contribution of scalar sea and the coefficients $b_{1}, b_{8}$ and $d_1, d_8$ come with the contribution of vector and tensor sea respectively. These coefficients ($a_{0}, a_{8}, a_{10}, b_{1}, b_{8}, b_{10}, c_{8}, d_{8}$) are associated with each state in the wavefunction has their own importance. They contain information about the various properties of hadrons like masses, magnetic moment, spin distribution etc. The details of all the terms of above-mentioned wave function can be found in refs. \cite{34,35}.
\section{Quadrupole Moment and Charge radii}
Charge radii ($r_B^{2}$) and quadrupole moments ($Q_B$) are the lowest order moments of the charge density $\rho$ in a low-momentum expansion. They are used to characterize the total charge, spatial extension, and shape of the baryons. Here, intrinsic quadrupole moment means the one obtained in a body-fixed-coordinate system that rotates with the nucleon. The nucleonic vertex function in terms of Electromagnetic Dirac and Pauli form factors $F_{1}(Q^{2})$ and  $F_{2}(Q^{2})$ is represented as:
\begin{equation}
\Gamma^{\mu}=F_{1}(Q^{2})\gamma^{\mu}+\kappa F_{2}(Q^{2}(i \frac{\sigma^{\mu\nu}q_{\nu}}{2m}))
\end{equation}
where $\kappa$ is the anomalous part of the magnetic moment, $\gamma^{\mu}$ are Dirac matrices and $\sigma^{\mu\nu} = i(\gamma_\mu\gamma_\nu- \gamma_\nu\gamma_\mu)/2$. Further, the electric and magnetic Sach form factors $G_{E}(Q^{2})$ and $G_{M}(Q^{2})$ \cite{36} can be related as:
\begin{equation}
G_{E}= F_{1}- \tau\kappa F_{2}
\end{equation}
\begin{equation}
G_{M}=F_{1}+\kappa F_{2}    
\end{equation}
where $\kappa =(\frac{Q}{2m})^{2}$.  The study of electromagnetic form factors of hadrons has received a lot of attention over the past years. The Fourier transform of the elastic form factors provide insights about the radial variation of the charge $\rho(r)$ and current j(r) densities \cite{37}. The observation by the physicists suggests that the ratio between electric and magnetic form factors decreases sharply for $1<Q^{2}< 6 GeV^{2}$ \cite{38}. The experimental evidence shows that there is a variation of charge distribution from spherical symmetry. It was suggested earlier that the quadrupole moment of nucleon should vanish on account of spin-$\frac{1}{2}$ nature which is the subject of theoretical and experimental activity. In addition, quadrupole moments of the decuplet baryons are still not clear. The transition amplitude in the $\gamma +p\rightarrow\Delta^{+}$ gives us information about photon absorption amplitudes, the magnetic dipole $G_{M1}$ and electric quadrupole moment from $G_{E2}$ and $G_{C2}$ \cite{39} respectively. Electric quadrupole $(E_{2})$ matrix elements, including quadrupole moments provide a principal measure of nuclear deformation, rotation, and related collective structure. The recent data from the collaborations \cite{41} like Jlab, Mainz predicts a non-zero value of quadrupole moment. The ratio of electric quadrupole to the magnetic dipole amplitude i.e. $\frac{E_{2}}{M_{1}}$= $-0.025\pm 0.005$. Similarly, the measurements made by LEGS and Mainz collaboration \cite{26,42} conclude that nucleon and $\Delta^{+}$ baryons have a deformity from spherical symmetric distribution. The geometrical shape of baryons can be measured from intrinsic quadrupole moments \cite{43}:
\begin{equation}
    Q_{0}^{p}=\int d^{3}r \rho^{p}{(r)(3z^{2}-r^{2})}
\end{equation}
where $\rho^{p}(r)$ is not necessarily spherically symmetric charge density of proton. If the charge density is concentrated along the z-direction (symmetry axis of the particle), the term proportional to $3z^{2}$ dominates, $Q_{0}$ is positive, and the particle is prolate. If the charge density is concentrated in the equatorial plane perpendicular to z, the term proportional to $r^{2}$ prevails, $Q_{0}$ is negative, and the particle is oblate. The most general form of the multipole expansion of the charge density $\rho$ in the spin-flavor space can be expressed as:
\begin{equation}
    \rho= A\sigma_{i=1}^{3} e_{i} I-B
\end{equation}
In order to find the quadrupole moment of baryon, a general QCD unitary operator and QCD eigen states $|B\rangle$ are defined explicitly in terms of quark and gluons. The quadrupole moment operator in terms of spin-flavor space can be expressed as \cite{44}:
\begin{equation}
 { \widehat{Q}_B = B\sum_{i\neq j}^3 e_i(3\sigma_{iz}\sigma_{jz}-\sigma_i.\sigma_j) + C\sum_{i\neq j\neq k}^ 3 e_i(3\sigma_{jz}\sigma_{kz}-\sigma_j.\sigma_k)} 
\end{equation}
The z-component of the Pauli spin (isospin) matrix $\sigma_i$ is denoted by $\sigma_{iz}$ and $e_i$ is the charge of the i-th quark where i = (u,d,s). The expanded form of the quadrupole moment operator of octet and decuplet baryons \cite{22} can be written as:\\
\begin{equation}
    \widehat{Q}_{1/2}= 3B\sum_{i\neq j} e_i\sigma_{iz}\sigma_{jz}+ 3C\sum_{i\neq j\neq k} e_i\sigma_{jz}\sigma_{kz}+(-3B+3C)\sum_{i} e_i \sigma_{iz}+ 3B\sum_{i} e_i
\end{equation}
\begin{equation}
     \widehat{Q}_{3/2}= 3B\sum_{i\neq j} e_i\sigma_{iz}\sigma_{jz}+ 3C\sum_{i\neq j\neq k} e_i\sigma_{jz}\sigma_{kz}+(-5B+5C)\sum_{i} e_i \sigma_{iz}+ (3B-6C)\sum_{i} e_i
\end{equation}
In the operator formalism \cite{18,45}, the charge radii operator can be expressed in terms of the sum of one-, two-, and three-quark contributions-\\
\begin{equation}
  \widehat{r}_B^{2} = A\sum_{i}e_i.1 + B\sum_{i\neq j} e_i \sigma_{i}\sigma_{j} + C\sum_{i\neq j\neq k}e_i \sigma_{j}\sigma_{k}
\end{equation}
Further, for octet and decuplet baryons, the expression can be written as:
\begin{equation}
\widehat{{r}}^{2}_{1/2} = (A-3B)\sum_{i}e_i + 3(B-C)\sum_{i} e_i\sigma_{iz}   
\end{equation}
\begin{equation}
   \widehat{{r}}^{2}_{3/2} = (A-3B+6C)\sum_{i}e_i + 5(B-C)\sum_{i} e_i\sigma_{iz}  
\end{equation}
Here $\widehat{{r}}^{2}_{1/2}$ and $\widehat{{r}}^{2}_{3/2}$represents the charge radii operator for spin- $\frac{1}{2}$ and spin-$\frac{3}{2}$ particles respectively. The unknown parameters A, B and C arise in the expansion of quadrupole moment and charge radii operators including the contribution of orbital and color space [16-18]. These constants (A, B and C) are determined from the experimental data on charge radii and quadrupole moments.\\
For a better understanding of the internal structure of baryons, an approach based on operator formalism is employed. In this formalism, a suitable operator which is associated with the properties of baryons, are applied to the terms of the wavefunction mentioned in equ. (1) $\&$ (2) represented as:
\begin{equation}
    \begin{aligned}
       \langle\Phi_{1/2,3/2}^{(\uparrow)}|{\widehat{O}}| \Phi_{1/2,3/2}^{(\uparrow)}\rangle =& \frac{1}{N^2} [{a_0}^{2}{\langle \Phi_{1}}^{(\frac{1}{2} \uparrow)}|\widehat{O}|{\Phi_{1}}^{(\frac{1}{2}\uparrow)}\rangle + {a_8}^2 \langle{\Phi_8}^{(\frac{1}{2}\uparrow)}|\widehat{O}|{\Phi_8}^{(\frac{1}{2}\uparrow)}\rangle + \\& {a_{10}}^2 \langle{\Phi_{10}}^{(\frac{1}{2}\uparrow)}|\widehat{O}|{\Phi_{10}}^{(\frac{1}{2}\uparrow)}\rangle + {b_1}^{2}\langle{\Phi_1}^{(\frac{1}{2}\uparrow)}|\widehat{O}|{\Phi_1}^{(\frac{1}{2}\uparrow)}\rangle + {b_8}^2 \langle{\Phi_8}^{(\frac{1}{2}\uparrow)}|\widehat{O}|{\Phi_8}^{(\frac{1}{2}\uparrow)}\rangle \\&+ {c_8}^2 \langle{\Phi_8}^{(\frac{3}{2}\uparrow)}|\widehat{O}|{\Phi_8}^{(\frac{3}{2}\uparrow)}\rangle + {d_8}^2 \langle{\Phi_8}^{(\frac{3}{2}\uparrow)} |\widehat{O}|{\Phi_8}^{(\frac{3}{2}\uparrow)}\rangle]    
    \end{aligned}
\end{equation}
where $\widehat{O}$ indicates the quadrupole moment and charge radii operator. Based on this formalism, we are able to derive certain expressions for the quadrupole moments of baryons listed in Table 1. Similarly, the obtained expression for charge radii of octet and decuplet particles are displayed in Table 2.
 \begin{table}[h!]
     \centering
     \small\addtolength{\tabcolsep}{-2pt}
     \begin{tabular}{|p{2 cm}|p{5.5 in}|}\hline
        Baryon & $\langle\Phi_{3/2}^{(\uparrow)}|Q_B^{2}|\Phi_{3/2}^{(\uparrow)}\rangle N^{2}$\\\hline
        p& ${a_0}^2$(-0.333B) + ${a_8}^2$(0.822B) + ${a_{10}}^2$(B) + ${b_1}^2$(0.9464B - 2.22C) +  ${b_8}^2$(0.176B - 0.296C) + ${b_{10}}^2$(0.0574B) + ${c_8}^2$(1.889B + 2.445C) + ${d_8}^2$(4.4B -0.06C)\\ \hline
        n& ${a_0}^2$(0.666B) + ${a_8}^2$(0.3779B) + ${b_1}^2$(-1.111B + 1.77C) + ${b_8}^2$(-0.341B + 0.126C) + ${c_8}^2$(-0.111B + 0.444C) + ${d_8}^2$(-0.466B + 0.7998C)\\ \hline
        $\Delta^{++}$ & ${a_0}^2$(2B +4C) + ${b_1}^2$(-3.523B + 8.121C) + ${b_8}^2$(-4.328B +7.856C) + ${d_1}^2$(-1.2B - 0.796C) + ${d_8}^2$(-1.894B -0.215C)\\\hline
        $\Delta^{+}$ & ${a_0}^2$(B -2C) + ${b_1}^2$(-1.66B + 1.838C) + ${b_8}^2$(-2.041B + 1.816C) +${d_1}^2$(0.192B - 1.704C) + ${d_8}^2$(0.491B - 2.314C)\\ \hline
        $\Delta^{0}$ & ${b_1}^2$(0.622B - 0.355C) + ${b_8}^2$(0.933B - 0.533C) + ${d_1}^2$(-0.533B + 1.066C) + ${d_8}^2$(-0.533B + 1.066C)\\ \hline
        $\Delta^{-}$ & ${a_0}^2$(-B - 2C) + ${b_1}^2$(1.762B - 4.06C) + ${b_8}^2$(2.164B - 3.928C) + ${d_1}^2$(0.066B + 0.133C) + ${d_8}^2$(0.893B + 0.214C) \\\hline
        \end{tabular}
        \caption{Expression obtained after applying quadrupole moment operator to the baryon octet $J^P = \frac{1}{2}^+$ and decuplet $J^P = \frac{3}{2}^+$ particles}
     \label{tab:my_label}
 \end{table}
 \begin{table}[h!]
     \centering
     \small\addtolength{\tabcolsep}{-2pt}
     \begin{tabular}{|p{1.5 cm}|p{5.5 in}|}\hline
Baryon & $\langle\Phi_{3/2}^{(\uparrow)}|r_B^{2}|\Phi_{3/2}^{(\uparrow)}\rangle N^{2}$\\\hline
 p& ${a_0}^2$(0.333A +0.666B -1.666C)+ ${a_8}^2$(0.333A +2.110B -3.110C)+ ${a_{10}}^2$(0.333A- B) +${b_1}^2$(0.647A -2.498B +0.555C) +${b_8}^2$(1.352A -4.613B +0.555C) +${b_{10}}^2$(0.646A - 1.940B) + ${c_8}^2$(0.549A +0.018B -1.666C) + ${d_8}^2$(0.066A- 1.266B +1.066C)\\ \hline
 n& ${a_0}^2$(-1.33B + 1.33C) + ${a_8}^2$(-0.178B +0.178C) + ${b_1}^2$(0.444B - 0.444C) + ${b_8}^2$(0.059B -0.059C) + ${c_8}^2$(-0.222B +0.222C) + ${d_8}^2$ (0.134B - 0.134C)\\ \hline
$\Delta^{++}$ & ${a_0}^2$(2A + 4B + 2C) + ${b_1}^2$(0.910A + 5.269B - 2.539C) + ${b_8}^2$(0.420A + 6.739B - 5.478C) + ${d_1}^2$(0.4A +0.8B+ 0.533C) + ${d_8}^2$(-0.647A +2.942B -4.885C)\\\hline
$\Delta^{+}$ & ${a_0}^2$(A+2B+C) + ${b_1}^2$(0.168A+3.601B -3.098C) + ${b_8}^2$(0.210A + 4.036B - 3.406C c) + ${d_1}^2$(0.198A +0.406B -0.212C) + ${d_8}^2$(0.533A - 1.099B + 2.698C)\\ \hline
$\Delta^{0}$ & ${b_1}^2$(-0.88B + 0.88C) + ${b_8}^2$ (-1.33B + 1.33C)\\ \hline
$\Delta^{-}$ & ${a_0}^2$(-A -2B - C) +${b_1}^2$ (-0.168A - 3.163B + 2.66C) + ${b_8}^2$(-0.210A - 2.869B + 2.239C) + ${d_1}^2$(-0.21A - 0.4B -0.2C) + ${d_8}^2$(-0.533A+1.099B -2.699C)\\\hline
\end{tabular}
\caption{Expressions obtained after applying charge radii operator to the baryon octet $J^P$= $\frac{1}{2}^+$ and decuplet $J^P$= $\frac{3}{2}^+$ particles}
     \label{tab:my_label}
 \end{table}
Our derived expressions involved two input parameters as listed in Table 1 and 2. One corresponds to the parameters used in operator formalism (A, B and C). The values of these parameters (A, B and C) are fitted by taking the available experimental data on quadrupole moment and charge radii as input. The other ones are statistical coefficients ($a_0, a_8, a_{10}, b_1, b_8, b_{10}, c_8, d_8$) that plays a pivotal role in determining the static properties of baryonic systems. As we know, there is a dynamic sea present with the valence quarks in the structure of baryons, where sea is a cluster consisting of gluons and quark-antiquarks pairs. The contribution of sea in spin, flavor and color space are determined in terms of these parameters ($a_0, a_8, a_{10}, b_1, b_8, b_{10}, c_8, d_8$). In order to calculate these parameters, we apply a statistical approach with the detailed balance principle. 
\section{Statistical model}
Hadrons are assumed to be an ensemble of valence quarks and sea containing various quark-gluon Fock states. These quarks and gluons in the Fock states are basically the "intrinsic" partons that are non-perturbatively multiconnected to the valence quarks. The complete set of quark-gluon Fock states can be expressed in expanded form as:
\begin{equation}
|Baryon \rangle = \sum_{i,j,k} C_{i,j,k} |(q^3),(i,j,k)\rangle
\end{equation}
Here $q^3$ represents the three valence quarks of the baryon, i is the number of quark-antiquark $u\bar{u}$ pairs, j is the quark-antiquark $d\bar{d}$ pairs and k is the number of gluons \cite{27}. The probability of finding the baryon in quark-gluon Fock state is $\rho_{i,j,k} = |C_{i,j,k}|^2$ and satisfy the condition of normalization: $\sum_{i,j,k} \rho_{i,j,k} = 1$.\\
The statistical model assumes the decomposition of baryonic state in various quark-gluon Fock states such as $|qqq\rangle|g\rangle, |qqq\rangle|gg\rangle$, $|qqq\rangle|q\bar q\rangle$ with the possibility of gluon splitting into $q\bar q$ pairs and the annihilation of quark-antiquark pairs into gluons. The octet and decuplet wavefunction of baryons includes the flavor, color, and spin space. To find the individual contribution of flavor, spin and color space, we calculate their relative probabilities. For the relative probability in flavor space, we depend upon the principle of detailed balance \cite{46}. This principle proposed by Zhang et al. \cite{47}, based on the assumption that every two subensembles balance with each other in a way: \\ 
$$\rho_{i,j,k} |\{q^{3}\},\{i,j,k\}\rangle \xLongleftrightarrow{balance}\rho_{i',j',k'} |\{q^{3}\},\{i',j',k'\}\rangle$$\\
The basic assumption of balancing any two ensembles is that the probability of finding hadrons in any Fock state should not change over time. The “go out” probability balance the “come in” probability for any Fock state.\\
\centerline{$\rho_A R_{A\rightarrow B} = \rho_B R_{B\rightarrow A}$}\\
Here, A and B represent the complete set of all Fock states. $\rho_A$ is the probability of finding the hadrons in Fock state A, $\rho_B$ is the probability of finding the hadrons in Fock state B and $R_{A\rightarrow B}$, $R_{B\rightarrow A}$ is the transition probability of Fock State from A to B and B to A. Various transition subprocesses such as q $\Leftrightarrow$ qg, g $\Leftrightarrow q\bar q$ and g $\Leftrightarrow$ gg are considered to calculate the transition probability of different Fock states in flavor space can be found in ref. \cite{29,47}. The transition probability talks about the chances of gluon(s) $\Leftrightarrow$  $u\bar u$, $d\bar d$, $s\bar s$ or $u\bar u \Leftrightarrow d\bar d$ states etc. This approach will change the Fock state $\&$ hence change the probability. The transition probabilities $R_{A\rightarrow B}$, $R_{B\rightarrow A}$ work on the same principle. 
This principle has been successfully explain the flavor asymmetry of hadrons \cite{27}, Parton distribution functions \cite{48}. Moreover, the probabilities in spin and color space are calculated from the multiplicities in the statistical model. 
The multiplicities for all Fock states are computed in the form of $\rho_{p,q}$ where the relative probability for valence part has 'Spin p' and sea carries 'Spin q' such that the resultant spin should come out as 1/2 for octet baryons and 3/2 for decuplet baryons. Similarly, the probabilities for the color spaces can be calculated which yields the color singlet state. 
Detailed calculations of these multiplicities are discussed in ref. \cite{27}. Further, the product of probabilities in spin and color space can be written in terms of a common multiplier “c” computed in the form of “nc”, where n is the multiplicity factor for each Fock state \cite{29}. The sum of the total probabilities in spin and color space will give the coefficients $a_0, a_8, a_{10}, b_1, b_8, b_{10}, c_8, d_8$ to the total wavefunction.  
These coefficients give us the contribution of sea quarks in computing the different properties of baryons. The main part within the statistical model which fascinates us that all the properties of baryons such as masses, spin distribution, magnetic moment etc. are interrelated with these probabilities (flavor, spin and color). Detailed information for the calculation of all the parameters using the statistical model is given in Refs. \cite{27,35}. The statistical model in combination with the principle of detailed balance is reliable for describing the properties like flavor asymmetry, semi-leptonic decays, masses, magnetic moments, spin distribution and axial coupling ratios of baryon octet and decuplet particles \cite{03, 28,29,30,31}.
\section{Results and Discussion}
Low-energy electromagnetic properties of baryons like charge radii and quadrupole moments are very important to probe their internal structure. 
These properties are computed by taking the effect of scalar (spin 0), tensor (spin 2) and vector (spin 1) sea. Individual contribution suggests the dominancy of sea (scalar, vector $\&$ tensor) to the quadrupole moment and charge radii of baryons. To examine the separate contribution of sea, we assume $a_{0,8,10} \neq 0$ and $b_{1,8,10}, c_8, d_8 = 0$  for scalar sea; whereas for vector sea $b_{1,8,10}, c_8 \neq 0$ and $a_{0,8,10}, d_8 = 0$; for tensor sea contribution, $d_8 \neq 0$ and $b_{1,8,10}, c_8, a_{0,8,10} = 0$ for octet baryons. In similar manner, the individual contribution of sea (scalar, vector and tensor) can be obtained for decuplet particles. It is worthy to mention that the statistical coefficients ($a_{0,8,10}, b_{1,8,10}, c_8, d_{1,8}$) demonstrate the importance of scalar, vector and tensor sea for determining the properties (charge radii and quadrupole moment) of baryons. These parameters are used as an input in the present work which is taken from the ref. \cite{27,35}. Different forms (Model C, P and D) of the statistical model help to find the role of sea dynamics to several static properties. We analyzed the properties (charge radii, quadrupole moment) by using Model D which is stated by Singh $\&$ Upadhyay \cite{27} that the contribution of states with higher multiplicities is suppressed. The suppression is based on the assumption that probability of a system in spin and color state is inversely proportional to the multiplicity (both in spin and color spaces) of the state. We would also like to mention that by taking the available experimental data on charge radii and quadrupole moment as input, we fit the parameters A, B and C by using $\chi^2$ minimization. The set of parameters (A, B and C) obtained after fitting are presented in Table 3 and 4.
\subsection{Charge radii}
To study the internal structure of hadrons, we are concerned with the two major crisis 'proton spin' and 'proton charge radius' which still remain unresolved. The proton radius puzzle has motivated new theoretical and experimental work. Recently PRad, an electron-proton scattering experiment at Jefferson Lab \cite{49}, reported a proton charge radius of $0.831\pm 0.007_{stat.} \pm 0.012_{sys.}$ fm. Moreover, in the three most recent H-spectroscopy measurements [50-52], two experiments found a small radius ($r_p=0.833\pm0.010$ fm) [50,52] and another one supports a larger value of charge radii ($r_p = 0.877$ fm) of proton \cite{51}. The electric charge radii are related to the spatial distribution of charged constituents (quarks) and provide information about the size of the hadrons. Using the statistical method, we present the calculated numerical values of electric charge radii of nucleons ($J^P=\frac{1}{2}^+$) and $\Delta$-hyperons ($J^P=\frac{3}{2}^+$) in Table 3. 
In order to get the significant contribution of sea, the statistical parameters ($a_{0,8,10}, b_{1,8,10}, c_8, d_{1,8}$) combine with valence part and the value of charge radii of baryons are determined. For nucleons (p, n), the dominancy of vector sea can be easily observed. Since the sea part is dominated by the emission of virtual gluons, we can expect the major contribution coming from the vector sea parameters ($b_1, b_8, c_8$) to the charge radii of nucleons. 
For $J^P=\frac{3}{2}^+$ particles, we observed that the scalar sea is an active contributor from the total sea. This might be due to the larger multiplicities of valence quark spin states when combined with the sea quarks spin (i.e. spin- 0,1 and 2). 
The probability of having spin-$\frac{3}{2}$ is more when coupled with spin-0 (scalar sea) as compared to spin-1 (vector sea). 
For both octet and decuplet baryons, the tensor sea appears to be less dominating due to the quark spin-flip process but its effect cannot be neglected. For $\Delta^0$, only vector sea contributed while pure scalar and tensor sea impact is zero. Because the contribution of valence quark across the scalar sea ($a_0$) and tensor sea ($d_1,d_8$) term is zero.\\ 
The importance of sea lies in the suppression of higher multiplicities of the quark-gluon Fock states. However in case, when the sea is excluded in the statistical model, the charge radii deviates more than $50\%$ 
from the computed value. The deviation is maximum for the proton and $\Delta^{++}$ is evident in Table 3.
On the contrary, if the total sea (scalar, vector, and tensor) is taken, the charge radii of nucleons is pretty close to the experimental value \cite{07}. On comparison our predicted value of charge radii shows good agreement in sign and magnitude with different phenomenological models [18, 21-23, 53-55]. 
Various theoretical models \cite{18,21} show the same sign for both particles $\Delta^+, \Delta^-$. Although, the difference in sign of charge radii of $\Delta^+, \Delta^-$ matches with the Lattice QCD predictions \cite{56} within the error of 20$\%$. Also, the three parameters A, B and C are essential for valence quark distribution in charge radii. In order to understand the individual contribution of one-, two- and three-quark term, we calculated the charge radii corresponding to one- and two-quark terms by taking the value of C=0. After putting C=0, the results are affected, as charge radii decreases in case of nucleons and for decuplet baryons, predicted values are increased upto 40$\%$. It indicates that a small variation in 'C' leads to a considerable change in charge radii values.\\
 
\begin{table*}[h!]{\normalsize
\renewcommand{\arraystretch}{1.0}
\renewcommand{\tabcolsep}{0.15cm}
    \centering
    \footnotesize{
    \begin{tabular}{cccccccccc}\toprule\toprule
   & & \multicolumn{6}{c}{Statistical Model with SU(3) symmetry} &  &\\\cmidrule{3-8}
   Baryons & \Longunderstack{Charge radii}&\Longunderstack{With\\A=0.754\\B=0.062\\C=-0.337}&\Longunderstack{Scalar\\Sea}&\Longunderstack{Vector\\Sea} &\Longunderstack{Tensor\\Sea}&\Longunderstack{Without\\Sea}&\Longunderstack{ With\\A=0.754\\B=0.062\\C=0}&Ref.[22]& Ref.[56] \\\midrule
     p & 0.756A -1.576B -0.693C & 0.707 &0.352 &0.402 &-0.020 & 4.126 & 0.472 & 0.766 & 0.685\\
      n & -0.292B +0.292C & -0.116&-0.02&-0.084 & 0.002&-0.436 & -0.018& -0.116& -0.158\\
     $\Delta$$^{++}$ & 1.909A + 3.862B+ 1.871C & 1.049&1.018 &0.028&0.0 &4.409&1.681 & 0.996 &-\\
     $\Delta$$^{+}$ & 0.951A + 1.940B + 0.894C & 0.537 &0.509& 0.028&0.0 &3.165& 0.839& 0.983 & 0.410\\
     $\Delta$$^{0}$ & -0.0106B + 0.0106C & -0.004 &0.0 &-0.004 &0.0&-0.882&0.0& -0.025 &-\\
     $\Delta$$^{-}$ & -0.965A - 1.951B - 0.943C & -0.531 &-0.518&-0.0103 &0.0 & -2.393&-0.849 & 1.033 & -0.410\\ \bottomrule\bottomrule
    \end{tabular}
 }}
 \caption{Numerical values for the charge radii of nucleons and spin-$(\frac{3}{2})^{+}$ decuplet baryons in the units of [fm$^2$]}
 \label{tab:my_label}
\end{table*} 
\subsection{Quadrupole moment }
Quadrupole moments are significant for providing evidence for the non-sphericity of baryons. They are important for determining the shape of baryons. Within the statistical approach, we present the computed values of quadrupole moments in Table 4. The individual contributions of scalar, vector and tensor sea are also presented in the Table 4. For nucleons (p, n), vector sea is the only active contributor from the total sea. This means that the quadrupole moment is mainly affected by vectorial sea parameters ($b_1, b_8, c_8$) having spin 1. For decuplet baryons ($\Delta^{++}, \Delta^+, \Delta^-$), the scalar sea contribution is greater as compared to vector and tensor sea. Due to the quark spin-flip process, the tensor sea has negligible contribution. The valence quark coupled with spin 0,1 and 2 (scalar, vector and tensor sea respectively) resulting into a large no. of spin states. The estimated multiplicities from spin states suggests that the higher multiplicity have a lower probability of survival. This might indicate that the chances of having spin-$\frac{3}{2}^+$ is more with the scalar sea (spin-0) as compared to vector and tensor sea. In case of $\Delta^0$, neither vector nor scalar sea contributes, but a little bit of contribution comes from the tensor sea.\\
In general, when the sea is completely excluded in the statistical model, the value of quadrupole moment shows a large deviation 
from the computed value which shows the active participation of sea quarks. A considerable change occurs only in the magnitude of the quadrupole moment but not in the sign. From this, we concluded that the impact of sea quarks do not directly affect the shape of baryons but rather it influences the overall structure of baryons. The sea contribution in quadrupole moment are given in Table 4 from different models. As there is no experimental information available for the quadrupole moments of baryons, we compare our findings with other theoretical predictions \cite{16,55,57,58} and they are consistent with the ref. \cite{21,23}. In the statistical model, the value of quadrupole moment observed an oblate shape for p, n, $\Delta^{++},\Delta^0$ baryons while a prolate shape for $\Delta^+, \Delta^-$ baryon. 
The quadrupole moment of $\Delta^+, \Delta^-$ show strange behavior with respect to the parameter 'C' which includes the contribution of three-quark term. If we put C=0, a -ve sign for $\Delta^+$ and a +ve sign for $\Delta^-$ observed which is consistent with other theoretical models \cite{16,21,53}. Although, a considerable change in the value of quadrupole moment observed after putting C=0. This shows the importance of three-quark term parameter 'C' to the quadrupole moment values. Similarly, by putting B=0, the contribution corresponding to the three-quark term can be calculated. The results are presented in the table below:
\begin{table*}[h!]{\normalsize
\renewcommand{\arraystretch}{1.0}
\tabcolsep 0.1cm
    \centering
   \small{
   \begin{tabular}{cccccccccc}\toprule\hline
           &           & \multicolumn{6}{c}{Statistical model with SU(3) symmetry} &  &\\\cmidrule{3-8}
   Baryons & \Longunderstack{Quadrupole\\ moment} & \Longunderstack{With\\B=-0.006\\C=-0.094} & \Longunderstack{Scalar\\Sea} & \Longunderstack{Vector\\Sea}&\Longunderstack{Tensor\\Sea} & \Longunderstack{Without\\Sea} &\Longunderstack{With\\B=-0.006\\C=0}& Ref.[21,23] & Ref.[16] \\\midrule
   p & 0.707B + 0.292C & -0.032 & 0.0 &-0.030 &-0.001 &-0.043&-0.004& -0.032 &-\\
      n & 0.4128B + 0.0241C & -0.038 &-0.001 &-0.034 & -0.002&-0.326&-0.055& -0.019 &-\\
     $\Delta$$^{++}$ & 1.842B + 3.863C & -0.375&-0.372 & -0.006&0.002 &-1.884 &-0.011& -0.343 & -0.12 \\
     $\Delta$$^{+}$ & 0.929B -1.939C & 0.177 &0.171 &-0.002& 0.007 &0.064& -0.005& -0.171 & -0.06\\
     $\Delta$$^{0}$ & 0.018B + 0.046C & -0.004 &0.0 &0.00&-0.004 &-0.120&0.0&0.0 & 0.0\\
     $\Delta$$^{-}$ & -0.939B - 1.940C & 0.188 &0.186 &0.003& 0.0& 0.885&0.005& 0.171 & 0.06\\ \bottomrule \bottomrule
    \end{tabular}
    }}
     \caption{Numerical values for the quadrupole moments of nucleons and spin-$(\frac{3}{2})^{+}$ decuplet baryons in the units of [fm$^2$]}
    \label{tab:my_label}
\end{table*}

\section{Conclusion}
In the present paper, a statistical approach with the principle of detailed balance is used to calculate the electromagnetic properties i.e. charge radii and quadrupole moments of the nucleons and $\Delta$-hyperons. Our main purpose is to study the impact of sea quarks on the properties of baryons in the framework of statistical model with detailed balance principle. It is important to study the sea because we expect the sea to be  a decisive factor for determine the properties of baryons. A suitable anti-symmetric wavefunction for baryons written in spin, color, flavor space is operated with charge radii and quadrupole moment operator. 
Since the sea contains quark-gluon Fock states, the statistical principle determine the probabilities for these Fock states in terms of statistical parameters ($a_{0}, a_{8}, a_{10}, b_{1}, b_{8}, b_{10}, c_{8}, d_{8}$). With the help of these parameters, we deeply explore the individual sea contribution to the charge radii and quadrupole moment of baryons. The parameter 'C' considerably contributes to the spin-spin interaction term of three-quarks. 
The quadrupole moment and charge radii are associated with the spin contribution and spatial charge distribution of the particles respectively. And the sea spin contribution is reflected in the results of these properties. 
Based on our analysis of the quadrupole moment, we conclude that the sea quarks have a pronounced impact on the overall structure of baryons, without directly inducing any changes in their shape. Also, our numerical analysis appreciated the presence of sea (scalar, vector and tensor) and gives crucial information about the geometrical structure of baryons. It is important to mention that our calculations performed in a non-relativistic frame and hold good for the hadronic energy scale of order 1 GeV$^2$. 
\section{Acknowlegement}
The authors gratefully acknowledge the financial support by the Department of Science and Technology (SERB/F/9119/2020), New Delhi.\\
\appendix
\section{Appendix 1}
    The expansion of quadrupole moment operator are given below:\\
\begin{equation}
 \widehat{Q}_B = B\sum_{i\neq j}^3 e_i(3\sigma_{iz}\sigma_{jz}-\bold{\sigma_i.\sigma_j}) + C\sum_{i\neq j\neq k}^ 3 e_i(3\sigma_{jz}\sigma_{kz}-\bold{\sigma_j.\sigma_k})
\end{equation}
\begin{equation}
  \widehat{Q}_B = B\left[\sum_{i\neq j}^3 3e_i\sigma_{iz}\sigma_{jz}-\sum_{i\neq j}^3 e_i(\bold{\sigma_{i}.\sigma_{j})}\right] + C\left[\sum_{i\neq j\neq k}^ 3 3e_i\sigma_{jz}\sigma_{kz}-\sum_{i\neq j\neq k}^3 e_i\bold{(\sigma_j.\sigma_k)}\right]  
\end{equation}

The spin operator terms included in above eq. can be written as [23]:\\
\begin{equation}
 \sum_{i\neq j}e_i\bold{(\sigma_i.\sigma_j)}= 2J\sum_{i}e_i\sigma_{iz}-3\sum_{i}e_i\\
 \end{equation}
\begin{equation}
    \sum_{i\neq j\neq k}e_i\bold{(\sigma_j.\sigma_k})= \pm3\sum_{i}e_i-\sum_{i \neq j}e_i\bold{(\sigma_i.\sigma_j)}  
\end{equation} 
In equ. (19), + sign holds for J=$\frac{3}{2}$ states and the - sign holds for J=$\frac{1}{2}$ states. After putting these operator values, equ. (17) becomes:
\begin{equation}
 \widehat{Q}_B= B\sum_{i\neq j}^3 3e_i\sigma_{iz}\sigma_{jz}-B\sum_{i\neq j}^3\left(2J\sum_{i}e_i\sigma_{iz}+3\sum_{i}e_i\right) + C\sum_{i\neq j\neq k}^ 3 3e_i\sigma_{jz}\sigma_{kz}-C\sum_{i\neq j\neq k}^ 3\left(\pm3\sum_{i}e_i-\sum_{i \neq j}e_i\bold{\sigma_i.\sigma_j}\right)   
\end{equation}
Using the expectation value of operator $2J\sum_{i}e_i\sigma_{iz}$, the operator in equ. (18) and (19) for octet and decuplet states written as: \\

For $\bold{J=\frac{1}{2}}$ \hspace{8cm}For $\bold{J=\frac{3}{2}}$\\
$\sum_{i\neq j}e_i\bold{(\sigma_i.\sigma_j)}=3\sum_{i}e_i\sigma_{iz}-3\sum_{i}e_i$ \hspace{3cm}$\sum_{i\neq j}e_i\bold{(\sigma_i.\sigma_j)}=5\sum_{i}e_i\sigma_{iz}-3\sum_{i}e_i$\\

 $\sum_{i\neq j\neq k}e_i\bold{(\sigma_j.\sigma_k})= -3\sum_{i}e_i\sigma_{iz}$\hspace{3cm}$\sum_{i\neq j\neq k}e_i\bold{(\sigma_j.\sigma_k})= 6\sum_{i}e_i-5\sum_{i}e_i\sigma_{iz}$\\
 
After substituting the above values in equ. (20), the expression for the quadrupole moment operator of octet and baryons can be expressed as:\\
\begin{equation}
    \widehat{Q}_{1/2}= 3B\sum_{i\neq j} e_i\sigma_{iz}\sigma_{jz}+ 3C\sum_{i\neq j\neq k} e_i\sigma_{jz}\sigma_{kz}+(-3B+3C)\sum_{i} e_i \sigma_{iz}+ 3B\sum_{i} e_i
\end{equation}
\begin{equation}
     \widehat{Q}_{3/2}= 3B\sum_{i\neq j} e_i\sigma_{iz}\sigma_{jz}+ 3C\sum_{i\neq j\neq k} e_i\sigma_{jz}\sigma_{kz}+(-5B+5C)\sum_{i} e_i \sigma_{iz}+ (3B-6C)\sum_{i} e_i
\end{equation}\\
The expansion of charge radii operator given as:\\
\begin{equation}
  \widehat{r}_B^{2} = A\sum_{i}e_i.1 + B\sum_{i\neq j} e_i \sigma_{i}\sigma_{j} + C\sum_{i\neq j\neq k}e_i \sigma_{j}\sigma_{k}
\end{equation}
Substituting the operator value of equ. (18) and (19) in equ. (23), the expression for charge radii of octet and decuplet baryons obtained:
\begin{equation}
\widehat{{r}}^{2}_{1/2} = (A-3B)\sum_{i}e_i + 3(B-C)\sum_{i} e_i\sigma_{iz}   
\end{equation}
\begin{equation}
   \widehat{{r}}^{2}_{3/2} = (A-3B+6C)\sum_{i}e_i + 5(B-C)\sum_{i} e_i\sigma_{iz}  
\end{equation}


\begin{thebibliography}{}
\bibitem{01} C. A. Aidala, S. D. Bass, D. Hasch, and G. K. Mallot, Rev. Mod. Phys. 85, 655 (2013).
 \bibitem{02} Xiaotong Song, Int. Journal of Modern Phys.A, Vol. 16, 22, pp. 3673-3697 (2001); M. Batra, A. Upadhyay, Nucl.Phys.A 889, pp. 18-28 (2012).
\bibitem{03} M. Batra, A. Upadhyay, Nucl.Phys.A 922, pp. 126-139 (2014).
\bibitem{04} H. Singh, A. Kumar, H. Dahiya, Eur.Phys.J.A 54, 7, 120 (2018).
\bibitem{05} S. Taylor, G. S. Mutchler, G. Adams et al., Phys. Rev. C, vol. 71, 5, 054609 (2005).
\bibitem{06} I. Eschrich, H. Krügeri, J. Simon et al., Phys. Lett. B, vol. 522, no. 3-4, pp. 233–239 (2001); F. He, C.R. Ji, W. Melnitchouk, A.W. Thomas, P. Wang, Phys.Rev.D 106, 5, 054006 (2022).
\bibitem{07}  R.L. Workman et al. (Particle Data Group), Prog. Theor. Exp. Phys. 2022, 083C01 (2022); E. Shuryak, I. Zahed, Phys.Rev.D 107,3, 034026 (2023).
\bibitem{08} C.B. Crawford, A. Sindile, T. Akdogan et al., Phys. Rev. Lett., vol. 98, no. 5, 052301 (2007); M. Gough Eschrich (Heirg, Max Planck Inst.) et al., SELEX Collaboration (2001);
Published in: Phys.Lett.B 522 (2001) 233-239 • e-Print R. Pohl, R. Gilman, G. A. Miller, K. Pachucki, Ann. Rev. Nucl. Part. Sci. 63, pp. 175–204 (2013); H. Atac, Z.-E Meziani, Nature Commun. 12, 1, 1759 (2021).
\bibitem{09} M.I. Krivoruchenko and M.M. Giannini, Phys. Rev. D 43, 3763 (1991).
\bibitem{10} R.K. Sahoo, A.R. Panda, and A. Nath, Phys. Rev. D 52, 4099 (1995).
\bibitem{11} G. Ramalho and K. Tsushima, Phys. Rev. D 87, 093011 (2013).
\bibitem{12} Yong-Lu Liu, Ming-Qiu Huang, Dao-Wei Wang, Eur.Phys.J.C 60, pp. 593-601 (2009).
\bibitem{13} K. Azizi, Eur. Phys. J. C 61, 311 (2009).
\bibitem{14} J. Kunz and P.J. Mulders, Phys. Rev. D, vol. 41, 5, pp. 1578-1585 (1990); C. Gobbi, S. Boffi and D.O. Riska, Nucl. Phys. A, vol. 547, 4, pp. 633-644 (1992).
\bibitem{15} A. J. Buchmann and E. M. Henley, Phys. Rev. C 63, 015202 (2001).
\bibitem{16} A. J. Buchmann and E. M. Henley, Phys. Rev. D 65, 073017 (2002).
\bibitem{17}  R. F. Lebed and D. R. Martin, Phys. Rev. D 70, 016008 (2004). E. E. Jenkins, Phys. Rev. D 85, 065007 (2012); A.J. Buchmann, J. A. Hester, and R. F. Lebed, Phys. Rev. D 66, 056002 (2002).
\bibitem{18} A. J. Buchmann and R. F. Lebed, Phys. Rev. D 67, 016002 (2003).
\bibitem{19} C. Alexandrou, G. Koutsou, H. Neff, J. W. Negele, W. Schroers, and A. Tsapalis, Phys. Rev. D 77, 085012 (2008); D. Arndt and B.C. Tiburzi, Phys. Rev. D 68, 114503 (2003); S. Boinepalli et al, Phys. Rev. D 80, 054505 (2009).
\bibitem{20} C. Alexandrou, G. Koutsou, J. W. Negele, Y. Proestos, and A. Tsapalis, Phys. Rev. D 83, 014501 (2011).
\bibitem{21} N. Sharma and H. Dahiya, Adv. High Energy Phys. 2013, 756847 (2013).
\bibitem{22} N. Sharma and H. Dahiya, Pramana 80, 237 (2013).
\bibitem{23} H. Dahiya, Neetika Sharma, PoS LC,056 (2010). 
\bibitem{24} G Blanpied et al, Phys. Rev. Lett. 79, 4337 (1997); R. Beck et al, Phys. Rev. Lett. 78, 606 (1997).
\bibitem{25}  L. Tiator, D. Drechsel, S. S. Kamalov, and S. N. Yang, European Phys. J. A, vol. 17, no. 3, pp. 357–363 (2003).
\bibitem{26}  G. Blanpied, M. Blecher, A. Caracappa et al., Phys. Rev. C, vol. 64, no. 2, 025203 (2001).
\bibitem{27} J.P. singh, A. Upadhyay, J. Phys. G, 30, pp. 881-894 (2004).
\bibitem{28} A. Kaur, A. Upadhyay, Eur. Phys. J. A 52, 11, 332 (2016); J. Zhang,B. Zhang, B. Q. Ma, Phys. Lett. B 524, 260 (2001).
\bibitem{29} M. Batra, A. Upadhyay, Int. J. Mod. Phys.A 28, 1350062 (2013).
\bibitem{30} A. Kaur, A. Upadhyay, Eur.Phys.J.A 52, 4, 105 (2016).
\bibitem{31} M. Batra, A. Upadhyay, Nucl. Phys. A 889, pp. 18-28 (2012).
\bibitem{32}  E. Golowich, E. Haqq, G. Karl, Phys. Rev. D 2, 160 (1983); F.E. Close, Z. Li, Phys. Rev. D 42, 2194 (1990).
\bibitem{33} F.E. Close, Rep. Prog. Phys. 51, 833 (1988); Z. Li, Phys. Rev. D 44, 2841 (1991).
\bibitem{34} X. Song, V. Gupta, Phys. Rev. D. 49, 2211 (1994).
\bibitem{35} A. kaur, A. Upadhyay, http://hdl.handle.net/10266/6081, (2018).
\bibitem{36} M. K. Jones et al., Phys. Rev. Lett. 84, 1398 (2000).
\bibitem{37} Alfons J. Buchmann, Few Body Syst. 59, 6, 145 (2018).
\bibitem{38} J. Arrington, Phys.Rev.C 68, 034325 (2003).
\bibitem{39} C. Beccho and G. Morpurgo, Phys. Lett., vol. 17, 3, pp. 352-354 (1965).
\bibitem{40}  R. F. Casten, Nuclear Structure from a Simple Perspective, 2nd ed., Oxford Studies in Nuclear Physics No. 23
(Oxford, 2000); D. J. Rowe, Nuclear Collective Motion: Models and
Theory (World Scientific, Singapore, 2010); D. J. Rowe and J. L. Wood, Fundamentals of Nuclear Models: Foundational Models (World Scientific, Singapore, 2010).
\bibitem{41} S. Taylor, G.S. Mutchler, G. Adams et al., Phys. Rev. C, vol. 71, 054609 (2005); O. Gayou, K.A. Aniol, T. Averett et al., Phys. Rev. Lett., vol. 88, 092301, (2002); C.B. Crawford, A. Sindile, T. Akdogan et al., Phys. Rev. Lett., vol. 98, 5, 052301 (2007); A.M. Bernstein, The European Phys. J.A., vol. 17, 3, pp. 349-355, (2003); C.N. Papanicolas, Eur. Phys. J.A., vol. 18, no. 2-3, pp. 141-145 (2003). 
\bibitem{42} R. Beck, H.P. Krahn, J. Ahrens et al., Phys. Rev. Lett., vol. 78, 4, pp. 606-609, (1997); L. Tiator, D. Drechsel, S.S. Kamalov and S.N. Yang, The European Phys. J.A., vol. 17, 3, pp. 357-363, (2003).  
\bibitem{43} J. M. Eisenberg and W. Greiner, Nuclear models (1970); P. Brix, Z. Naturforsch 41a, 3 (1986); P. Brix and H. Kopfermann, Z. Phys. 126, 344 (1949).
\bibitem{44} G. Morpurgo, Phys. Rev. D 40, 2997 (1989); G. Dillon and G. Morpurgo, Phys. Lett. B 448, 107(1999); A.J. Buchmann and E.M. Henly, Phys. Rev. D 65, 073017 (2002).
\bibitem{45} A.J. Buchmann, R.F. Lebed, Phys.Rev.D 62 ,96005 (2000).
\bibitem{46} Y.J. Zhang, B. Zhang, B.Q. Ma, Phys. Lett. B 523, 260 (2001); Y.J. Zhang, B.Q. Ma, L. Yang, Int. J. Mod. Phys. A 18, 1465 (2003).
\bibitem{47} Y.J. Zhang, Wei-Zhen Deng, B.Q. Ma, Phys. Rev. D 65, 114005 (2002).
\bibitem{48} Y.J. Zhang, et al., Phys. Lett. B, 528, pp. 228-232 (2002).
\bibitem{49}  W. Xiong et al., Nature 575, no. 7781, 147 (2019).
\bibitem{50} Beyer, A. et al., “The Rydberg constant and proton size from atomic hydrogen,” Science 358, 79–86 (2017).
\bibitem{51}  Fleurbaey, H. et al., Phys. Rev. Lett. 120, 183001 (2018).
\bibitem{52}  Bezginov N. et al., “A measurement of the atomic hydrogen Lamb shift and the proton charge radius,” Science 365,
1007-1012 (2019).
\bibitem{53} B. Kubis and U.G. Meissner, The Eur. Phys.J. C, vol. 18, 4, pp. 747-756 (2001); D. Arndt and B.C. Tiburzi, Phys. Rev. D, vol. 68, 9, 094501 (2003).
\bibitem{54} P. Wang, D.B. Leinweber, Phys. Rev. D, vol.79, 9, 094001 (2009); D.B. Leinweber, T. Draper and R.M . Woloshyn, Phys. REv. D, vol. 46, 7, pp. 3067-3085 (1992).
\bibitem{55} S.J. Puglia, M.J.R. Musolf and S.L. Zhu, Phys. Rev. D, vol. 36, 034014 (2001); K. Berger, R.F. Wagebrunn and W. Plessas, Phys. Rev. D, vol. 70, 9, 094927 (2004).
\bibitem{56} B. Schwesinger and H. Weigel, Nucl. Phys. A, vol. 540, no. 3-4, pp. 461–477, (1992); J. Kroll and B. Schwesinger,  Phys. Lett. B, vol. 334, pp. 287–289 (1994).
\bibitem{57} M. N. Butler, M. J. Savage, and R. P. Springer, Phys. Rev. D, vol. 49, pp. 3459–3465, (1994); M. K. Banerjee and J. Milana,Phys. Rev. D, vol. 54, no. 9, pp. 5804–5811 (1996).
\bibitem{58} M. I. Krivoruchenko and M. M. Giannini, Phys. Rev. D, vol. 43, 11, pp. 3763–3765 (1991).
\end{thebibliography}
\end{document}